\documentclass[conference]{IEEEtran}
\IEEEoverridecommandlockouts
\usepackage{cite}
\usepackage{amsmath,amssymb,amsfonts}
\usepackage{algorithmic}
\usepackage{graphicx}
\usepackage{textcomp}
\usepackage{xcolor}
\usepackage{url}

\usepackage{pifont}
\usepackage[ruled,linesnumbered]{algorithm2e}
\usepackage{multirow}
\usepackage{multicol}
\usepackage{threeparttable}
\usepackage{grumble}
\usepackage{makecell}
\usepackage{tabularx}
\usepackage[caption=false,font=footnotesize,labelfont=sf,textfont=sf]{subfig}

\usepackage{hyperref}
\hypersetup{
hidelinks,
colorlinks=false,
linkcolor=false,
citecolor=false,
urlcolor = false
}

\def\BibTeX{{\rm B\kern-.05em{\sc i\kern-.025em b}\kern-.08em
    T\kern-.1667em\lower.7ex\hbox{E}\kern-.125emX}}
\begin{document}

\title{
  Memory-Efficient and Secure DNN Inference on TrustZone-enabled Consumer IoT Devices
}


\author{Xueshuo~Xie$^{2, \dagger }$,
        Haoxu~Wang$^{1, \dagger}$,
        Zhaolong~Jian$^{1}$,
        Tao~Li$^{1,2, *}$,
         Wei~Wang$^{4}$,
        Zhiwei Xu$^{2}$, and
        Guiling~Wang$^{3}$
        \\ 
\textit{$^1 $ College of Computer Science, Nankai University, China}\\
\textit{$^2 $ Haihe Lab of ITAI, China}\\
\textit{$^3 $ New Jersey Institute of Technology, USA}\\
\textit{$^4 $ Beijing Key Laboratory of Security and Privacy in Intelligent Transportation, Beijing Jiaotong University, China}

\IEEEcompsocitemizethanks{\IEEEcompsocthanksitem 
\IEEEcompsocthanksitem $^\dagger$ These authors contributed equally to this work.
\IEEEcompsocthanksitem $^*$ 
Corresponding author: Tao Li (litao@nankai.edu.cn).
}
}

\maketitle

\begin{abstract}

Edge intelligence enables resource-demanding Deep Neural Network (DNN) inference without transferring original data, addressing concerns about data privacy in consumer Internet of Things (IoT) devices. For privacy-sensitive applications, deploying models in hardware-isolated trusted execution environments (TEEs) becomes essential. However, the limited secure memory in TEEs poses challenges for deploying DNN inference, and alternative techniques like model partitioning and offloading introduce performance degradation and security issues.
In this paper, we present a novel approach for advanced model deployment in TrustZone that ensures comprehensive privacy preservation during model inference. We design a memory-efficient management method to support memory-demanding inference in TEEs. By adjusting the memory priority, we effectively mitigate memory leakage risks and memory overlap conflicts, resulting in 32 lines of code alterations in the trusted operating system.
Additionally, we leverage two tiny libraries: S-Tinylib (2,538 LoCs), a tiny deep learning library, and Tinylibm (827 LoCs), a tiny math library, to support efficient inference in TEEs. We implemented a prototype on Raspberry Pi 3B+ and evaluated it using three well-known lightweight DNN models. The experimental results demonstrate that our design significantly improves inference speed by 3.13 times and reduces power consumption by over 66.5\% compared to non-memory optimization method in TEEs.

\end{abstract}

\begin{IEEEkeywords}
Secure DNN Inference, Memory Management, ARM TrustZone, Tiny Deep Learning Framework.
\end{IEEEkeywords}

\section{Introduction}

\IEEEPARstart{A}{ccording} to Statista, the number of consumer  Internet of Things (IoT) devices is expected to reach 75 billion by 2025\cite{Statista_IOT}, with even larger numbers anticipated in the future. With the increasing processing capabilities of consumer IoT devices, local device computing has garnered significant attention due to its potential to reduce transmission delays and safeguard user privacy\cite{zhou2019edge}. However, models deployed on resource-constrained IoT devices still face serious risks of unauthorized or malicious user behaviors, necessitating robust security protection\cite{zhang2023understanding}. Trusted execution environments (TEEs), such as ARM TrustZone, offer secure inference without extensive ciphertext transformations, noisy data disturbances, or model function fitting, making them an attractive alternative to costly privacy-preserving techniques like differential privacy and homomorphic encryption, which may incur computation overhead and impact inference accuracy.

Despite the advantages of TEE-based secure inference, devices equipped with TEEs face two significant challenges. First, the available secure memory in TEEs, designed to protect highly sensitive information such as fingerprints and keys, is too limited to support memory-demanding DNN inference tasks. This limitation results in either inadequate protection or significant performance degradation\cite{jia2022hyperenclave}. Second, the Trusted-OS on TrustZone is constrained by the minimization design principle of the trusted computing base (TCB) and supports only a few underlying dependency libraries. Existing deep learning frameworks, often resource-demanding with complex dependencies, cannot be directly used within TrustZone. Additionally, the separation of DNN models and their computing libraries leads to compatibility issues\cite{liu2021trusted}.

This paper focuses on addressing these challenges by adaptively utilizing TEEs to provide comprehensive protection for pre-trained models and the entire inference process on consumer IoT devices. Our key design strategies revolve around two aspects. First, we propose Smart-Zone, a novel memory management solution that dynamically adjusts the secure memory size required by each pre-trained DNN model. Smart-Zone resizes the number of page tables in virtual memory and the secure memory region size in physical memory. Additionally, it re-allocates an optimal shared memory size using an exponential fitting function to efficiently transmit model parameters. To mitigate security and privacy concerns, we adjust memory region priorities to avoid the risk of data leakage due to memory overlap conflicts between shared memory and secure memory.

The second aspect of our design tackles supporting model inference in TrustZone while ensuring compatibility with various deep learning frameworks and minimizing the TCB. To achieve this, we introduce a tiny library named Tinylib, consisting of N-Tinylib, S-Tinylib, and Tinylibm. These libraries facilitate DNN inferences, with N-Tinylib operating in the Rich Execution Environment (REE) for pre-processing, S-Tinylib running in TEE to perform secure inference and post-processing, and Tinylibm providing efficient mathematical function support in the TEE. While maintaining a small TCB footprint, S-Tinylib and Tinylibm support a wide range of pre-trained DNN models and essential functionalities for common inference tasks on consumer IoT devices. Furthermore, we optimize the page table remapping operation of secure monitor call (\emph{SMC}) and computation flow in our solution to reduce interaction delay, incorporate vectorization algorithms for improved code parallel execution, and leverage Tinylibm to reduce inference time and power consumption.

We implemented a prototype on Raspberry Pi 3B+ to validate functionality and evaluate performance. The code size of S-Tinylib is 2,538 LoCs, and Tinylibm is 827 LoCs. With only 3,365 LoCs added to TCB, our design enables efficient secure inference on consumer IoT devices. We also made modifications to the OPTEE OS (32 LoCs). Our performance evaluation demonstrates a significant reduction of power consumption by more than 66.5\% and an improvement in inference speed by 3.13 times compared to non-memory optimization method in TEEs.

Furthermore, we discuss the extensibility of our design, including an extension tool to support the transformation of DNN models trained by other frameworks for convenient deployment on Tinylib. Additionally, we consider integrating state-of-the-art cryptography technologies to protect the input and results of inference and mitigate data leakage risks.

The contributions of this paper are as follows:
\begin{itemize}
\item We propose an efficient memory management solution (Smart-Zone) to adapt secure memory size for each DNN model, dynamically optimizing inference latency and power consumption on resource-constrained devices.
\item We design a compact and extensible library (Tinylib) following the small TCB principle, enabling seamless integration on TrustZone-enabled consumer IoT devices. With an extension tool, Tinylib supports additional pre-trained models from various frameworks.
\item We optimize the page table remapping operation of \emph{SMC} and the computation flow in our solution to reduce interaction delay, use vectorization algorithms to improve code parallel execution and leverage the Tinylibm math library to reduce inference time and power consumption.
\end{itemize}

\section{Preliminaries}
\subsection{ARM TrustZone}

\begin{table}[ht]
\caption{Comparisons among Different TEEs}\label{fl}
\begin{center}
\resizebox{\linewidth}{!}{
\begin{tabular}{|c|c|c|c|c|}
\hline
Arch & Name &Enclave Type&Enclave Num&Mem Granu\\
\hline
\hline
\multirow{3}*{Intel} & SGX & Process & Unlimited  & Page\\ \cline{2-5}
& Scalable SGX & Process & Unlimited  & Page\\ \cline{2-5}
& TDX & VM & Limited  & Page\\
\hline
\multirow{4}*{AMD} & SEV & VM & 16/256 & Page\\ \cline{2-5}
& SEV-ES & VM & Limited  & Page\\ \cline{2-5}
& SEV-SNP & VM & Limited  & Page\\ \cline{2-5}
& HyperEnclave\cite{jia2022hyperenclave} & Process & Unlimited  & Region\\ 
\hline
\multirow{5}*{RISC-V} & Sanctum\cite{2016Sanctum} & Process & DRAM regions  & Region\\ \cline{2-5}
& TIMBER-V\cite{2019TIMBER} & VM & Unlimited  & Page\\ \cline{2-5}
& Keystone\cite{lee2020keystone} &  VM & PMPs  & Region\\ \cline{2-5}
& PENGLAI\cite{feng2021scalable} &  VM & Unlimited  & Page \\
\hline
Power & Power9 PEF & VM & Unlimited  & Region \\ 
\hline
\multirow{7}*{ARM} 
& Komodo\cite{2017Komodo} & Process & Unlimited  & Region \\ \cline{2-5}
& Sanctuary\cite{brasser2019sanctuary} & VM & Unlimited  & Region\\ \cline{2-5}
& ARM S-EL2 & VM & Unlimited  & Region \\ \cline{2-5}
& ARM CCA\cite{li2022design} & VM & Unlimited  & Page \\ \cline{2-5}
& TEEv\cite{li2019teev} & VM & Unlimited  & Region\\ \cline{2-5}
& TwinVisor\cite{li2021twinvisor} & VM & Unlimited  & Page \\\cline{2-5}
& TrustZone & VM & Unlimited  & Region\\ 
\hline
\end{tabular}}
\end{center}
\label{tab:TEEs}
\end{table}

ARM TrustZone is a hardware-based security technology that provides a trusted execution environment (TEE) alongside the regular execution environment (REE) on ARM-based processors.
Existing TEEs provide hardware-isolated execution environments at the process and virtual machine (VM) levels. As shown in TABLE \ref{tab:TEEs}, TEEs are gradually evolving to offer more flexible and larger secure memory solutions\cite{jauernig2020trusted}. However, when deployed on consumer IoT devices, existing TEEs face challenges due to limited extensibility and constrained resources. ARM introduced TrustZone in its ARMv6 architecture to enhance system security and improved it in ARMv7 and ARMv8, enabling chip-level protection and hardware isolation on consumer IoT devices\cite{pinto2019demystifying}. TrustZone-enabled ARM cores operate in two states: Secure and Non-Secure. In the Secure state, only code on the TEE side can execute in isolated memory space, while in the Non-Secure state, only code on the REE side can execute in regular memory space. TrustZone Address Space Controller (TZASC) divides address spaces into secure and non-secure ranges, denying REE access to secure memory space. The Trusted Computing Base (TCB) of TrustZone includes trusted firmware (192,114 LoCs), TEE OS (242,849 LoCs) and TA applications. The TCB minimizes its size for thorough code base inspection and verification.

\subsection{Deep Learning Framework}
With the rapid development of deep learning applications, academia and industry have launched multiple deep learning frameworks to easily implement a complex model from scratch. To accelerate and optimize this process,  these frameworks can provide many basic libraries to support various computing and model structures\cite{xie2022performance}, such as Caffe, Pytorch. 
However, deep learning frameworks have progressively become more complicated to accommodate vast and complex neural networks.
As shown in TABLE \ref{tab:Frameworks}, we compared state-of-the-art deep learning frameworks in terms of lines of code (LoCs), files, and sizes. The size of TensorFlow, which is one of the most popular frameworks, has reached $1.2$GB.
For device computing, the limited resources on IoT devices can not support such a huge framework. The complex framework will also lead to a huge TCB and increase the system's attack surface.
Although there are some lightweight frameworks like Darknet\cite{redmon2013darknet}, using deep learning frameworks on TrustZone still faces challenges due to the lack of underlying supporting libraries and limited security resources.
This makes it considerably difficult for DNN inference to obtain TEE protection on consumer IoT devices.

\begin{table}
 \caption{Comparisons among Different Deep Learning Frameworks.}\label{ls}
 \begin{center}
 \resizebox{\linewidth}{!}{
 \begin{tabular}{|l|c|c|c|c|c|c|}
 \hline
 \multirow{2}*{} & \multicolumn{3}{c}{\textbf{Implementation Code (LoCs)}} & & \multirow{2}*{\textbf{Files}} & \multirow{2}*{\textbf{Size}} 
 \\ \cline{2-5}
 & C & C++  & Python/Java  & \textbf{Totals} & & \\
 \hline
 \hline
 Caffe & / & 42,856  & 5,426  & 81,364 & 611 & 98MB 
 \\
 \hline
 Caffe2 & 2,115 & 49,747  & 48,133  & 172.248 & 1.207 & 264MB 
 \\
 \hline
 CNTK & / & 113,916  & 52,265  & 343,078 & 1,784 & 1.4GB 
 \\
 \hline
 DL4J & / & 225,006  & 539,148  & 1,038,770 & 7481 & 458MB 
 \\
 \hline
 MXNet & 6,203 & 103,771  & 117,956  & 462,404 & 2,587 & 234MB 
 \\
 \hline
 MindSpore & 97,198 & 799,545  & 419,166  & 1,703,220 & 17,900 & 1.7GB 
 \\
 \hline
 OneFlow & / & 153,059  & 89,799  & 338,774 & 3,462 & 74MB 
 \\
 \hline
 Paddle & 371 & 109,503  & 64,678  & 288,623 & 2,689 & 340MB 
 \\
 \hline
 PyTorch & 37,069 & 776,395  & 584,817  & 1,988,427 & 9,272 & 679MB
 \\
 \hline
 TensorFlow & 1,419 & 1,725,593  & 660,333  & 3,170,692 & 15,672 & 1.2GB 
 \\
 \hline
 Theano & 19,912 & /  & 129,556  & 158,059 & 481 & 170MB 
 \\
 \hline
 Darknet & 20375 & 604  & 256  & 31,438 & 141 & 5.24MB 
 \\
 \hline
 \end{tabular}}
 \end{center}
 \label{tab:Frameworks}
 \end{table}

\subsection{Secure DNN inference in TEEs}

When the pre-trained models are deployed on consumer IoT devices, applying TEEs to protect the integrity and legitimate usage of the models, and the secure inference processes, has been successful.
Slalom\cite{tramer2018slalom} uses TEEs and GPU collaboration for private model inference based on an efficient outsourcing scheme for matrix multiplication.
However, this method requires outsourcing many computing and storage resources and is difficult to deploy on consumer IoT devices. 
To overcome the limited computation capability and power provision on devices, 
Lasagna\cite{li2021lasagna} offloads inference tasks to the edge cloud and uses SGX to protect the model's data, but this brings significant network transmission delay.
Trusted-DNN\cite{liu2021trusted} transmits the weight of the model in segments, and each segment is calculated separately in TEE conforming to the limited secure memory resources of IoT devices; 
however, TEE cannot save all weight data of the model, resulting a complete data transmission for every image inference. This causes significant delay. 
DarkneTZ\cite{mo2020darknetz} runs the sensitive layers of the DNN model in the secure space of TrustZone on the edge devices through the model partition;
however, many layers are in the untrusted environment, resulting in inveitable leakage of data, likely intermediate data modification by attackers, and potential wrong inference results. 

\textbf{Conclusion:} 
The aforementioned solutions by model partition and task offloading have to compromise between model integrity and model inference efficiency on consumer IoT devices, and thus fail to solve the problem fundamentally.
Additionally, these solutions still retain a lot of unused libraries in the deep learning framework, resulting in large TCB.
These are precisely the problems that our work mainly solves.

\section{Smart-Zone}

\subsection{Desin Principle}
\label{sec:threatmodel}
\begin{figure}[ht]
     \centering
     \includegraphics[width=0.43\textwidth]{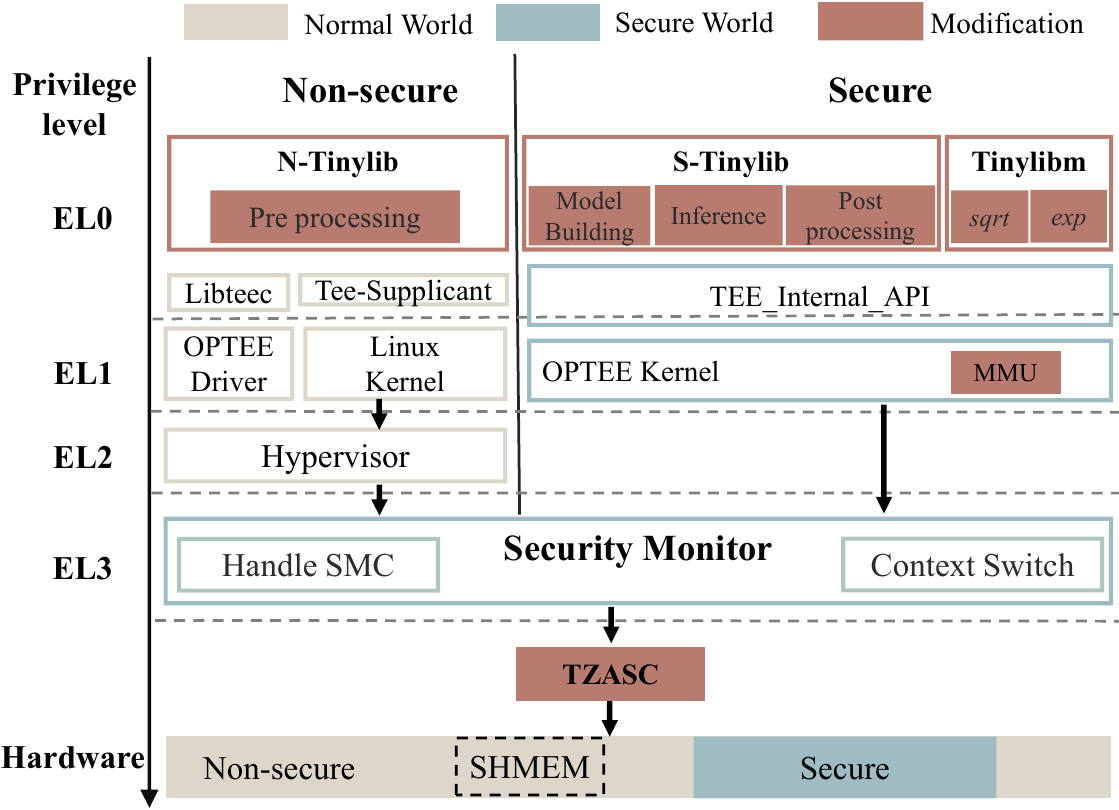}
     \caption{The Design Overview of Smart-Zone.}\label{fig:model}
\end{figure}

As shown in Fig. \ref{fig:model}, Smart-Zone aims to mitigate attacks on deploying pre-trained DNN models by protecting the entire model at a low cost in TrustZone-enabled consumer IoT devices. The primary compatibility consideration of Smart-Zone can run and integrate on common TrustZone-enabled consumer IoT devices.  We use standard TEE system architectures and corresponding APIs, such as OPTEE. We first present our design goals:
\begin{itemize}
     \item \textbf{Extensibility.} Smart-Zone supports the easy deployment of a variety of pre-trained DNN models from other frameworks and supports secure memory re-allocation based on the requirements of DNN models.
     \item \textbf{Performance.} Smart-Zone does not introduce large performance overhead on inference time and power consumption.
     \item \textbf{Security.} Smart-Zone achieves the above two goals with security guarantees, e.g., the integrity of pre-trained models and a trusted inference process.
\end{itemize}

By introducing a tiny library Tinylib, our design allows many pre-trained DNN models from diverse frameworks to effectively run on consumer IoT devices by only introducing a tiny expansion in TCB. The design brings two benefits on isolation and restricted interaction because we can completely protect the model inference using Smart-Zone through secure memory extension and optimization. 


\subsection{Threat Model.} 
We assume an adversary with full access to the REE on consumer IoT devices: this could be a real user, malicious third-party software installed on the devices, or a malicious, compromised OS. 
We only assume the TEE is trustable to guarantee data integrity and computing confidentiality.
We assume that a DNN model is pre-trained using private data from the server. 
The model providers can fully guarantee the model privacy during training on their servers by utilizing existing protection methods or even by training the model offline. 
The weight can be secretly provisioned to the user devices without other security concerns and privacy issues\cite{hussain2021coinn}. 
Each weight in the pre-trained model and the inference images are encrypted independently, and only the decryption engine in TEEs can decrypt it.   
We do not consider hardware-based attacks, such as hardware Trojan insertion\cite{chakraborty2013hardware}, fault injection\cite{chen2021voltpillager}, and side-channel analysis\cite{lou2021survey}. 
Their defense methods are orthogonal to our design. 
We consider the following security concerns:

\textbf{Authorized usage of the pre-trained models.} The valuable pre-trained model may be stolen for unauthorized use\cite{jiang2018secure}. 

\textbf{Tampering with the models.} 
The attackers may tamper with the model to jeopardize its integrity, or they may cause the model to incorrectly classify predefined inputs, such as through trojan attacks\cite{tang2020embarrassingly} and bit-flip attacks\cite{he2020defending, biggio2018wild}.

\textbf{Thwarting inference process.} 
An adversary can mislead and intentionally thwart the DNN model inference process by adding some noises or injecting fake data to the intermediate results of the model, e.g. fault injection attack\cite{liu2017fault}.

\subsection{On-demand Memory Management}
%
Physical memory is always divided into three parts for secure isolation on TrustZone-enabled consumer IoT devices: secure memory, shared memory, and normal memory. TEE uses a memory management unit (\emph{MMU}) to manage the secure memory space and establish the mapping between the physical address and virtual address, then uses the virtual address to access the data in physical memory and uses the page table to record the above mapping in different regions. As shown in Fig. \ref{fig:memory}, the virtual memory divides into different regions with different attributes and functions, such as memory region type, physical address, virtual address, size, and attributes. For example, region 6 is a secure memory region for Trusted Applications (TAs). 
When the CPU uses a virtual address reading or writing memory, the \emph{MMU} uses the page table to convert the virtual address to a physical address, and then sends the physical address and the secure read/write signal through \emph{AXI} bus. Only the operations with a secure read/write signal can access secure physical memory space. TEE uses \emph{TZASC} to divide secure and non-secure regions in physical memory, but the secure memory is always relatively small to only minimal requirements of TAs. For the above complex secure inference tasks, if we adjust memory regions blindly: (1) overlapping memory regions may lead to privacy data leakage; (2) unreasonable memory allocation will exacerbate the memory shortage problem of resource-constrained IoT devices, seriously reducing system performance.

\textbf{Secure Memory Extension.}
Our primary concern is setting a reasonable secure memory range of Smart-Zone on resource-constrained devices. We can calculate the pre-allocated memory space of each model layer and the memory size occupied by the model layer structure, and furthermore, accurately calculate the secure memory size required to deploy the DNN model to the Smart-Zone using the formula:
\begin{figure}[ht]
  \centering
  \includegraphics[width=0.45\textwidth]{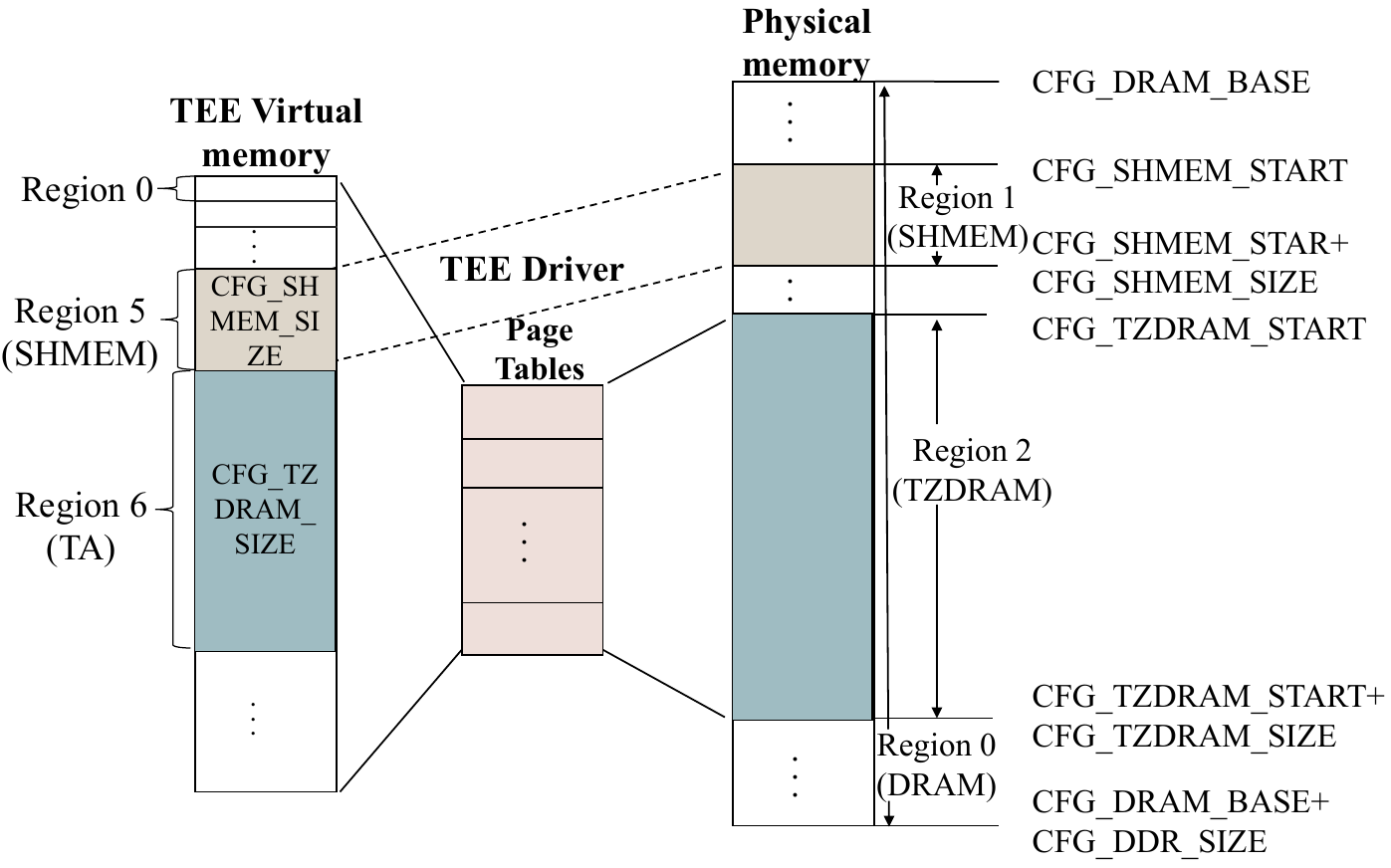}
  \caption{The mapping between virtual memory and physical memory through page table.}
  \label{fig:memory}
\end{figure}

\begin{equation}
TEE_{RAM} = M_s + F_{gc} + TEE_{CORE}
\label{eq1}
\end{equation}
where $M_s$ refers to the proportion of model parameters, $F_{gc}$ refers to the intermediate or final calculation results of the model in forwarding inference, and $TEE_{CORE}$ is the memory required by the TEE kernel. We evaluate some state-of-the-art DNN models that can entirely run on IoT devices by formula (\ref{eq1}), even lightweight DNN models require much more memory than the secure memory size provided by existing TEEs. For example, mobilenet requires 133.1MB, far exceeding the 16MB provided by OPTEE. 
When we obtain the value of $TEE_{RAM}$ by formula (1), we can re-allocate and extend the size of restricted secure memory, both in virtual memory and physical memory. For virtual memory, TEE OS uses page tables to limit the amount of virtual memory that can be used by the secure world. A page table corresponds to a \emph{MMU} page table entry. A \emph{MMU} page table entry in a 64-bit system can map 2MB memory.  
In a 64-bit system, we use $TEE_{RAM}$ to calculate the number of page tables by the following formula: 

\begin{equation}
NUM_{pgt}= \lfloor TEE_{RAM}/2 \rfloor +1
\label{eq3}
\end{equation}

We also extend the physical secure memory size of Smart-Zone, the secure memory region of mapped physical memory is configured by \emph{CFG\_TZDRAM\_START} and \emph{CFG\_TZDRAM\_SIZE} using \emph{TZASC} in Fig. \ref{fig:memory}. When virtual memory is mapped to physical memory, the secure memory remains the same size. We reset the \emph{CFG\_TZDRAM\_SIZE} equals $TEE_{RAM}$, to extend the secure range of physical memory. Thus, we can run and protect the entire pre-trained DNN model in the extended secure memory.

\textbf{Shared Memory Adjustment Optimization.}
Although we use Smart-Zone to fully protect the DNN model and reduce the interaction delay compared with the cross-layer solution, we still need to transfer data, such as weights, between REE and TEE through shared memory during the model building phase. If the shared memory is too small, the weight data needs to be transmitted multiple times, and each transmission requires a complete \emph{invoke} operation, which seriously reduces the weight transmission efficiency. But, excessive shared memory size will crowd out the available memory size of REE and TEE, resulting in a waste of memory. We design a dynamic weight transfer module to set the optimal shared memory size for each DNN model. We first perform the model's pre-processing process to obtain the DNN model's transmission delay with different shared memory sizes. 
We observe through experiments
the transmission delay is inversely proportional to the shared memory size, and use an exponential function to fit the relationship between the weight transfer delay and the shared memory size.
\begin{equation}
y = \alpha x^{\beta}
\label{eq4}
\end{equation}
where $x$ represents the shared memory size,  $y$ represents the transmission delay,
$ \alpha $ and $ \beta $ are fitting coefficients. 
When we get the appropriate shared memory size using cost function (\ref{eq4}), we modify $CFG\_SHMEM\_SIZE$ to set the shared memory size in \emph{TZASC} as shown in Fig. \ref{fig:memory}.

\textbf{Memory Overlap Conflict Resolution.}
As shown in Fig. \ref{fig:memory}, when we reset the size of shared memory $CFG\_SHMEM\_SIZE$, the end of the shared memory may overwrite the beginning of the secure memory, resulting in memory overlap conflicts. If the memory priority is not properly assigned, the conflict may cause the shared memory to overflow. Attackers may obtain the control permission of the secure memory, causing the risk of information leakage in the secure memory. 
For example, 
careless hardware manufacturers may inadvertently cause memory overlaps that are difficult to detect and exploited by attackers.
To solve the memory overlap conflict, the existing TEEs usually set the starting address of the shared memory region far from the starting address of the secure memory region, and the size of shared memory is very small for just exchanging little information.

In Smart-Zone, we optimize the memory priority assignment of \emph{TZASC}. Without the loss of generality, we also allocate the physical memory into three parts, but we set the secure memory priority to the highest. The address field of region 0 covers the entire memory space and can be accessed by REE. The address field of region 1 is a shared memory region and can be accessed by REE and TEE. The address field of region 2 is the secure memory space and can only be accessed by TEE. The size of region 2 should be equal to the extended secure memory size required by the DNN models. If we set the secure region priority to the highest, when other region address fields overlap with the secure region address field, the overlapping address space is still only accessed by TEE, thus avoiding privacy leakage. Of course, when we select the optimal shared memory size above, we can also determine whether there is an overlap with secure memory.

\begin{figure}[ht]
  \centering
  \includegraphics[width=0.43\textwidth]{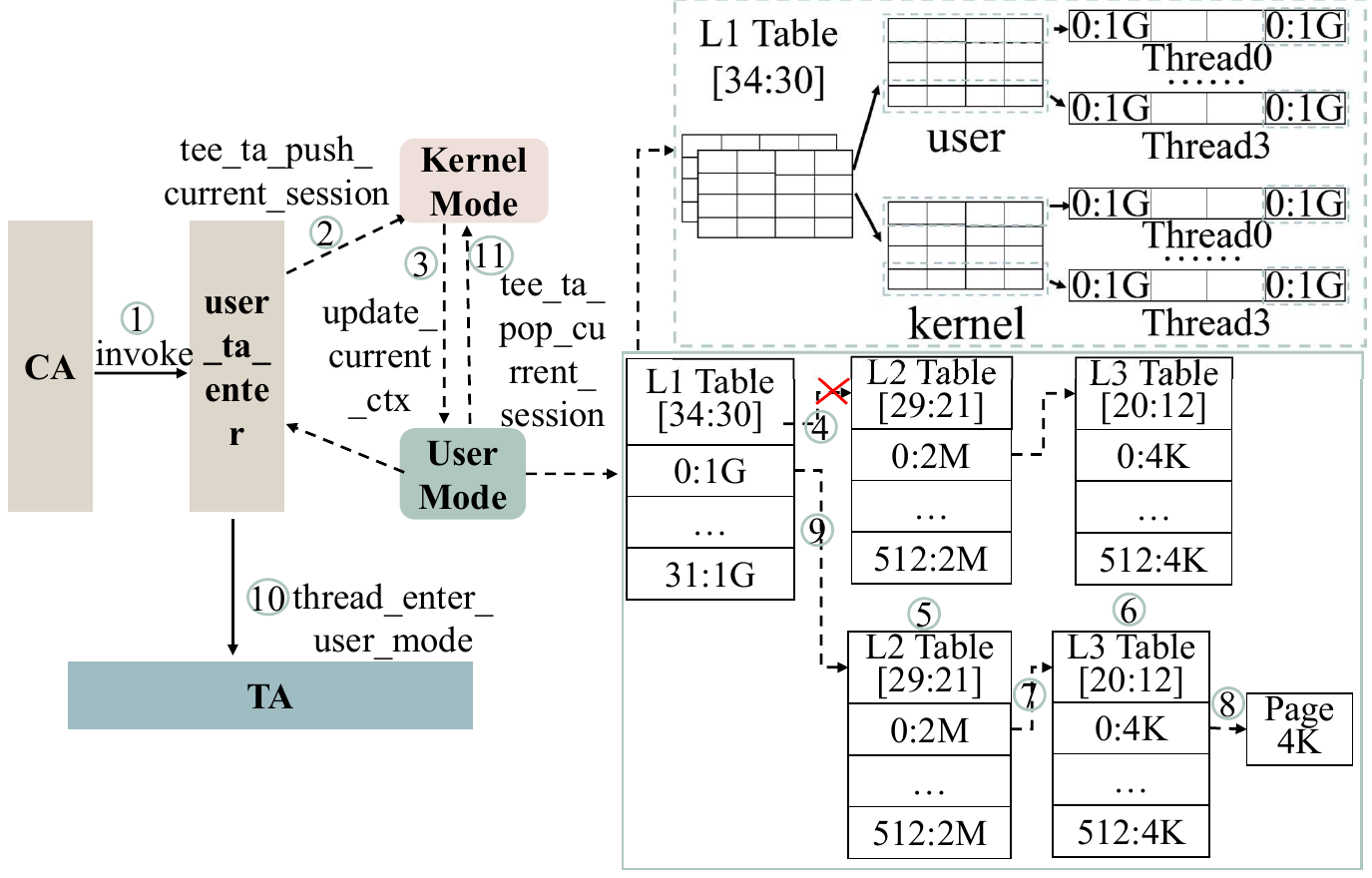}
  \caption{The \emph{invoke} operation flow optimization.}\label{fig:invoke}
\end{figure}

\textbf{Optimization of Extended Secure Memory.}
After extending the secure memory size in Smart-Zone, we observed the execution time of TAs increased to a certain extent. Because every time the TAs are invoked by the current thread, all allocated page tables will be released and reapplied, and then re-complete the mapping of three levels of page tables in total, as shown in Fig. \ref{fig:invoke}. 

The remapping operation can solve the deadlock caused by multiple threads competing for page tables when multiple threads are executing at the same time. When the size of secure memory is small, the remapping for \emph{invoke} operation will introduce negligible delays. But, 
the \emph{SMC} operation becomes a performance bottleneck in large extended secure memory of Smart-Zone.

In Smart-Zone, we will call the \emph{invoke} operation when we transmit data between REE and TEE. If we transmit weights in frequent interaction at the model building stage, we will often call the \emph{invoke} operation. This will cause the release and remapping of the page table and will result in a non-negligible delay. 
Due to the memory limitations of the end device and the high memory requirements of DNN inference, only one TA is usually running on the device at the same time.
Thus, there is no deadlock caused by multiple threads competing for pages. 
As shown in Fig. \ref{fig:invoke}, we recommend canceling each \emph{invoke}'s remapping of the page table where large memory is used and frequently invokes are required. More specifically, for a complete TA program, we complete the mapping of the page tables at \emph{TEEC\_Open\_Session} stage, and release the page tables at \emph{TEEC\_Close\_Session} stage. For each \emph{TEEC\_Invoke\_Command} operation, we perform TA-specific operations directly after completing the user-mode switch.

\subsection{Tinylib Supported DNN inference}
For secure DNN inference on TrustZone-enabled consumer IoT devices, we observed two limitations since the minimize TCB: (1) there is a lack of deep learning libraries to reuse more network layers and inference computing, resulting in repeated development when the different model is updated or upgraded;
(2) there is a lack of a math library similar to \emph{libm} in REE to support math calculations for model inference.

The convenient deployment of the pre-trained model requires a library which be easily integrated with TEE OS to implement the network layers, inference calculation, and math computing. 
Thus, we design a tiny library Tinylib with three parts: (1) \textbf{N-Tinylib} runs in REE, mainly realizing the functions of data loading and model pre-processing; 
(2) \textbf{S-Tinylib} runs in TEE and implements model building, inference calculation, and post-processing to minimize TCB, including the \emph{convolution} layer, \emph{pooling} layer, \emph{normalization}, \emph{activation} function, \emph{loss} function, \emph{im2col}, \emph{gemm} etc,; 
\begin{figure}[ht]
  \centering
  \includegraphics[width=0.42\textwidth]{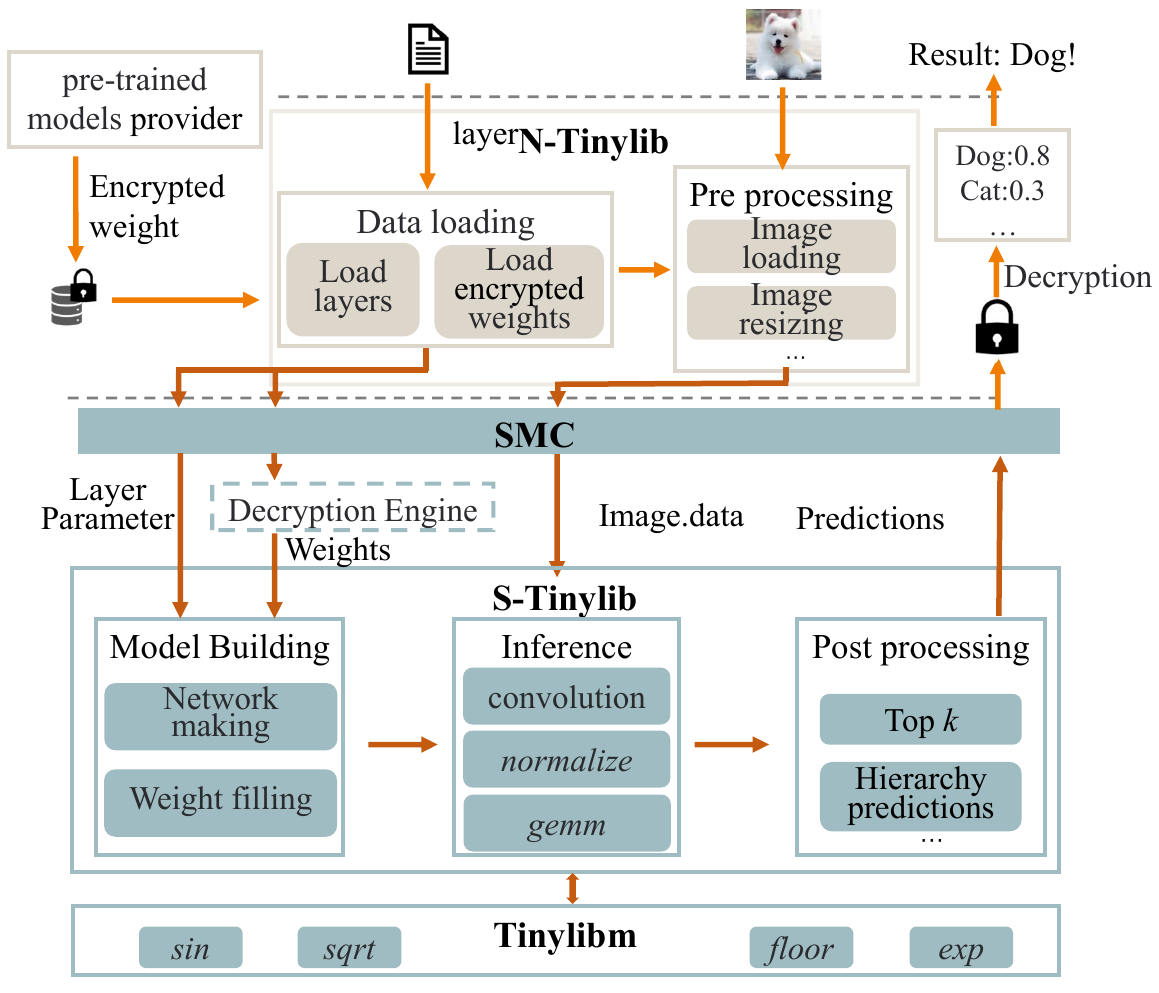}
  \caption{The inference flow chart of Tinylib.}
  \label{fig:tinylib}
\end{figure}
(3) \textbf{Tinylibm} is a tiny math library and runs in TEE to provide S-Tinylib with math calculation. Tinylibm just includes \emph{sqrt}, \emph{floor}, \emph{sin}, \emph{exp}, and other common math functions used in model inference. 
We compile Tinylibm with the \emph{IEEE754} floating-point arithmetic standard as a static function library in \emph{C} language, and it can be used directly in other TEEs environments. We also compile S-Tinylib into TEE OS, and Tinylib can easily run and integrate on common TrustZone-enabled consumer IoT devices. S-Tinylib and Tinylibm in TEE try to stay tiny as they minimize TCB but support as many features as one needs for common device secure inference tasks.

As shown in Fig. \ref{fig:tinylib}, in the inference process, N-Tinylib first reads the network layer configuration file of the model and transfers the network layer parameters to S-Tinylib by layer to build the model. S-Tinylib will request the memory space required by the model weight when building the network structure. Then, N-Tinylib reads the encrypted pre-trained model weight file and transfers the encrypted weight data from the REE to the decryption engine in the TEE by layer. The decryption engine transmits the decrypted weight data to S-Tinylib to fill in the weight data of the network layer. In this process, all the decrypted weight data are saved in TEE, so attackers cannot obtain any data of the pre-trained model. Next, N-Tinylib reads the input image for inference, performs resizing and regularizing operations on the image, stores the image information in pixels in the array, and uses the \emph{invoke} operation to transfer the array data to S-Tinylib to perform network layer calculations in TEE. 
After S-Tinylib performs post-processing, such as hierarchy prediction, S-Tinylib returns the encrypted result to the user application.
After decrypting the result, the application obtains the highest probability of $k$ objects and the corresponding object name.
When the inference of all images is completed, S-Tinylib deletes all network layer data, including the pre-trained model's weight. In this case, our design can extend to follow some mature secure exchange protocol, e.g. \emph{SIGMA}, and avoid the exchange data leakage risks and defense injecting fake data or adding noises to the input data.

\textbf{Compilation Optimization.} 
When deploying Tinylib to Smart-Zone, we will first compile the \emph{C} language codes to an executable file. \emph{GCC} provides some optimization options to improve the operation efficiency of Tinylib. For example, \emph{Ofast} compilation optimization can enable many vectorization algorithms to make full use of CPU pipeline, cache, etc., and improve the running speed of code. However, \emph{Ofast} compilation optimization is not perfect, may cause some running errors due to the violation of language standards, and increase the size of the executed code. Fortunately, we tested that enabling \emph{Ofast} will not affect the functional and inference accuracy of Tinylib in any way. Since the code size of S-Tinylib is very small, it will not significantly increase the compiled file size on this basis, but it can significantly improve the running speed.

\textbf{Computation Flow Optimization of Tinylib.}
We observed that when building the network structure, many deep learning frameworks first randomly generates a weight value for the \emph{convolution} layer and other weighted network layers, to discover good enough weights and make the different neurons more discriminative. 
However, the weight value of the pre-trained model will overwrite this random value during model inference, making the random value generation a waste. 
Considering this random weight generation takes a long time and increases inference delay during the model inference step, Tinylib directly read the pre-trained model parameters and covers this random weight value at the model building stage, and deletes the above process of random generation at the inference stage. The above optimization can greatly reduce the delay in model building.

\textbf{Extension Tool for Models Conversion.} We provide an extension tool to convert more pre-trained DNN models from other deep learning frameworks to deploy on Tinylib. Standalone Tinylib can support many common models, including ResNet, Darknet Reference,  YoLoV, etc. However, new models are constantly being proposed and the training of deep learning models is preferred to use some famous frameworks.
The tool first determines the training framework by reading the  input model file and then calls the corresponding conversion module to convert into a Caffe model by layer transition and weight transition separately. For example, in layer transition, the \emph{Pytorch2Caffe} module converts the \emph{Conv2d}, \emph{batchnorm2d} layer of Pytorch to the \emph{convolution} layer, \emph{Scale} layer and \emph{batchnorm} layer of Caffe; in weight transition, \emph{Pytorch2Caffe} module converts the \emph{weight} and \emph{bias} of \emph{Conv2d} into convolution data, and convert the \emph{weight}, \emph{bias}, \emph{running\_mean}, and \emph{running\_var} of \emph{batchnorm2d} into data of \emph{Scale} and \emph{batchnorm} in turn. Then, the above Caffe model is converted into Tinylib model by layer transition and weight transition separately. For example, in layer transition, \emph{Caffe2Tinylib} module converts the above \emph{convolution} layer, \emph{Scale} layer, and \emph{batchnorm} layer into Tinylib's \emph{convolutional} layer, and convert the above data of Caffe model to \emph{weights}, \emph{biases}, \emph{scales}, \emph{rolling\_mean}, and \emph{rolling\_variance} of Tinylib in turn.

\section{Evaluation}

\subsection{Experimental Setup}

\textbf{Prototype Implementation.} We implement a prototype on Raspberry Pi 3B+ with quad-core ARM Cortex-A53 (ARMv8) $64$-bit @ 1.4GHz CPU with 1GB LPDDR2 SDRAM0, and use OPTEE 3.8.0 as TEE and Linux 4.14 as REE. This prototype helps confirm that our design will function well on future commercial hardware with practicability and extensibility.

\begin{table}[ht]
\caption{Three state-of-the-art pre-trained Models.}\label{learning}
\begin{center}
\begin{tabular}{|c|c|c|c|c|}
\hline
 Model & Size & Layer & Parameters &  Mem Cost\\
\hline
\hline
Resnet-18\cite{he2016deep} & 44MB & 29 & 33,161,024 &  144.5MB \\ 
\hline
Darknet Reference & 28MB & 26 & 6,096,512 &  93.59MB \\ 
\hline
Mobilenet & 17MB & 30 & 5,347,520 &  133.1MB \\
\hline
\end{tabular}
\end{center}
\end{table}

\textbf{Models and Datasets.}
In TABLE \ref{learning}, we use three well-known lightweight pre-trained models (Resnet-18, Darknet Reference, and Mobilenet) on ImageNet for evaluation and use formula (\ref{eq1}) to calculate the memory consumption: $93.59$MB, $133.1$MB, and $144.5$MB, respectively. These models’ memory requirements far exceed the original TEE’s limits (around 16 MB), but can fully run in Smart-Zone.

\textbf{Implementation Complexity.} We run \emph{cloc}\footnote{https://github.com/uctakeoff/vscode-counter} tool to measure the code size of the prototype in TABLE \ref{tcb}. As shown in Fig. \ref{fig:model} and Fig. \ref{fig:tinylib}, only S-Tinylib and Tinylibm run in TEE and influence the TCB. For efficient math calculation, we choose Fdlibm as our baseline. The code size of Tinylibm is around $807$ LoCs, which is much smaller than the $5,203$ LoCs of Fdlibm. The number of files reduces from $82$ to $17$, and the compiled file size reduces from $252.3$KB to $40.2$KB, almost $80$\% reduction. For secure inference, we choose Darknet as our baseline. The code size of S-Tinylib is around $2,538$ LoCs, far less than $31,438$ LoCs of Darkent, almost $92$\% reduction. The files were also reduced from $141$ to $40$, the file size was reduced from $5.24$MB to $111.2$KB, and the compiled file size was reduced from $741.2$KB to $118.6$KB. We also have minor modifications on OPTEE and \emph{TZASC} ($32$ LoCs) to extend the secure memory size and optimize the inference performance. The small code size makes formal verification feasible and keeps a small TCB to reduce the attack surface.

\textbf{Measured metrics.}
We measured inference time (seconds), average power (watts), and power consumption (joule) three times and took the average to avoid the impact of environmental temperature and other factors.

We use Monsoon High Voltage Power Monitor\footnote{https://www.msoon.com/}, a high-precision power metering hardware capable with a voltage range of $0.8$V to $13.5$V and up to $6$A  current. We configured it to power the Raspberry Pi 3B+ using the required $5$V voltage while recording the power in a $5$KHZ sampling rate. In addition, we use the integral method to obtain the electrical energy consumed by the device over a period of time.

\begin{table}[ht]
\caption{The code size of the prototype.}\label{tcb}
\begin{center}
\begin{tabular}{|c|c|c|c|c|}
\hline
& LoCs & Files & Size & Complize Size \\
\hline
\hline
 Fdlibm & 5,203 & 82 & 274.4KB & 252.3KB \\
\hline
 \textbf{Tinylibm} & \textbf{827} & \textbf{17} & \textbf{65.9KB} & \textbf{40.2KB} \\
\hline
 Darknet & 31,438 & 141 & 5.24MB & 741.2KB \\ 
\hline
 Tinylib & 22041 & 155 & 934.2KB & 719KB\\
\hline
N-Tinylib & 18694 & 98 & 757.1KB &560.2KB \\
\hline
 \textbf{S-Tinylib} & \textbf{2,538} & \textbf{40} & \textbf{111.2KB} & \textbf{118.6KB} \\ 
\hline
OPTEE OS & 242,849 & 1979 & / & / \\
\hline
\textbf{OPTEE OS-Mod} & \textbf{32} & \textbf{7} & / & / \\
\hline
\end{tabular}
\end{center}
\end{table}

\begin{figure}[ht]
  \centering
  \includegraphics[width=0.33\textwidth]{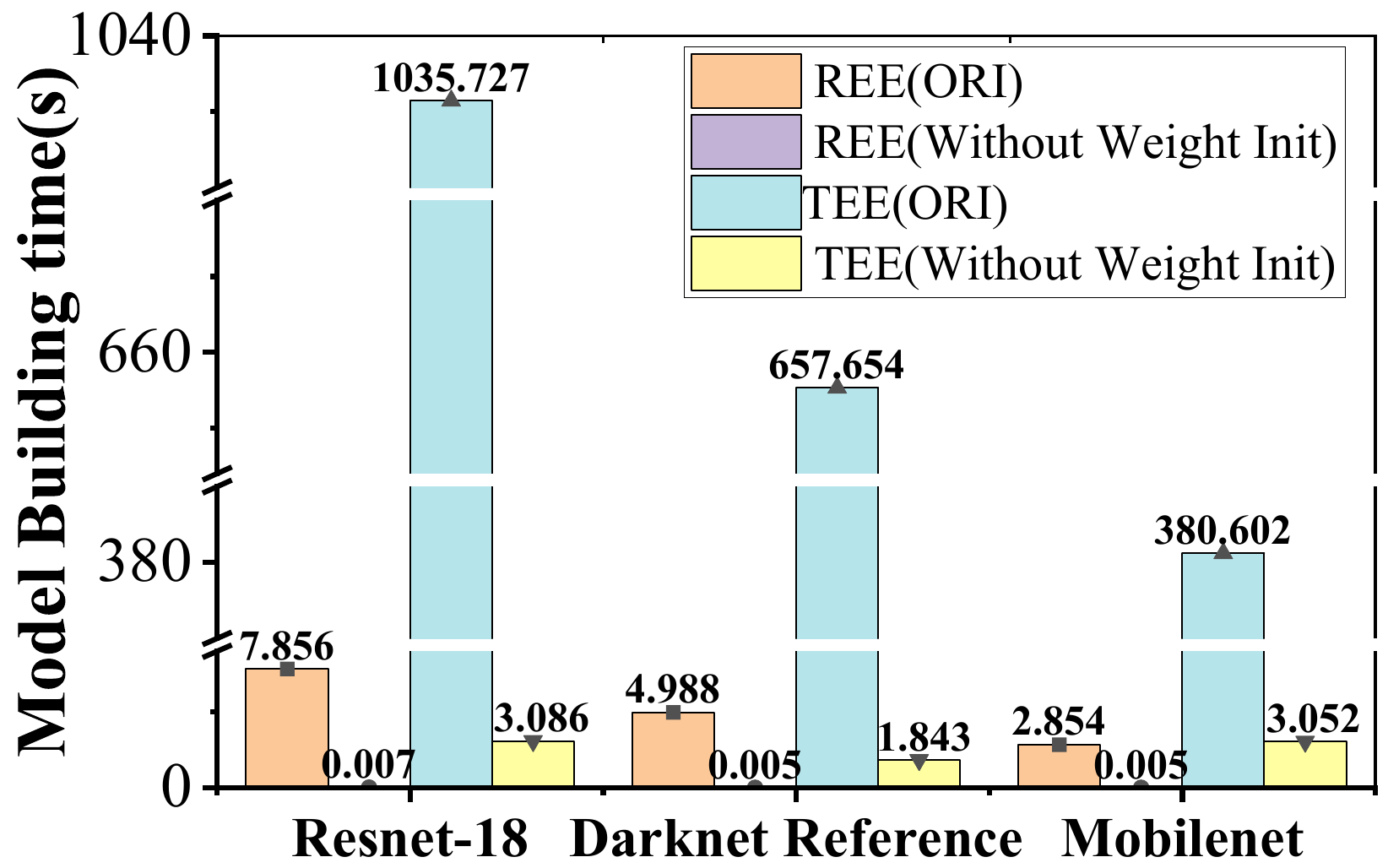}
  \caption{The performance on computation flow optimization of Tinylib at the model building stage.}\label{fig:build}
\end{figure}

\textbf{Baseline.}
We compare it against the following baselines:
Model Partition, DNN inference runs partly in REE and partly in TEE according to DarkneTZ\cite{mo2020darknetz}; REE(Without Weight Init) (or TEE(Without Weight Init)),  model building without initialization of random weights; REE (ORI), DNN inference all runs in REE; TEE (ORI), DNN inference all runs in TEE without any optimization; REE (Ofast) (or TEE (Ofast)), the \emph{Ofast} optimization is started on the basis of REE (ORI) (or TEE(ORI)); REE (Tinylib) (or TEE(Tinylib)), Tinylibm library is used on the basis of REE (Ofast) (or TEE (Ofast)).

\subsection{Performance Analysis}
\textbf{Inference time.} 
As shown in Fig. \ref{fig:build}, the time consumed by Resnet-18 is reduced from $1035.727$s to $3.086$s, almost $99.7$\% reduction in the TEE model building stage.
Darknet Reference and Mobilenet also decreased by $99.7$\% and $99.2$\% respectively.
This is because the process of calculating the random value of the weight in the model building stage requires a lot of math calculations, and the math library of DarkneTZ\cite{mo2020darknetz} used in TEE has lower performance. After our optimization, the model building delay in TEE decreased more markedly.
We also noticed that the model-building time in TEE is still higher than that in REE after optimization. This is because the building of each network layer in TEE requires a separate \emph{invoke} operation to pass parameters from CA to TA.

\begin{figure*}[ht]
\centering
\begin{minipage}[b]{0.26\linewidth}
\centering
\subfloat[][Resnet-18]{\includegraphics[width=1\textwidth]{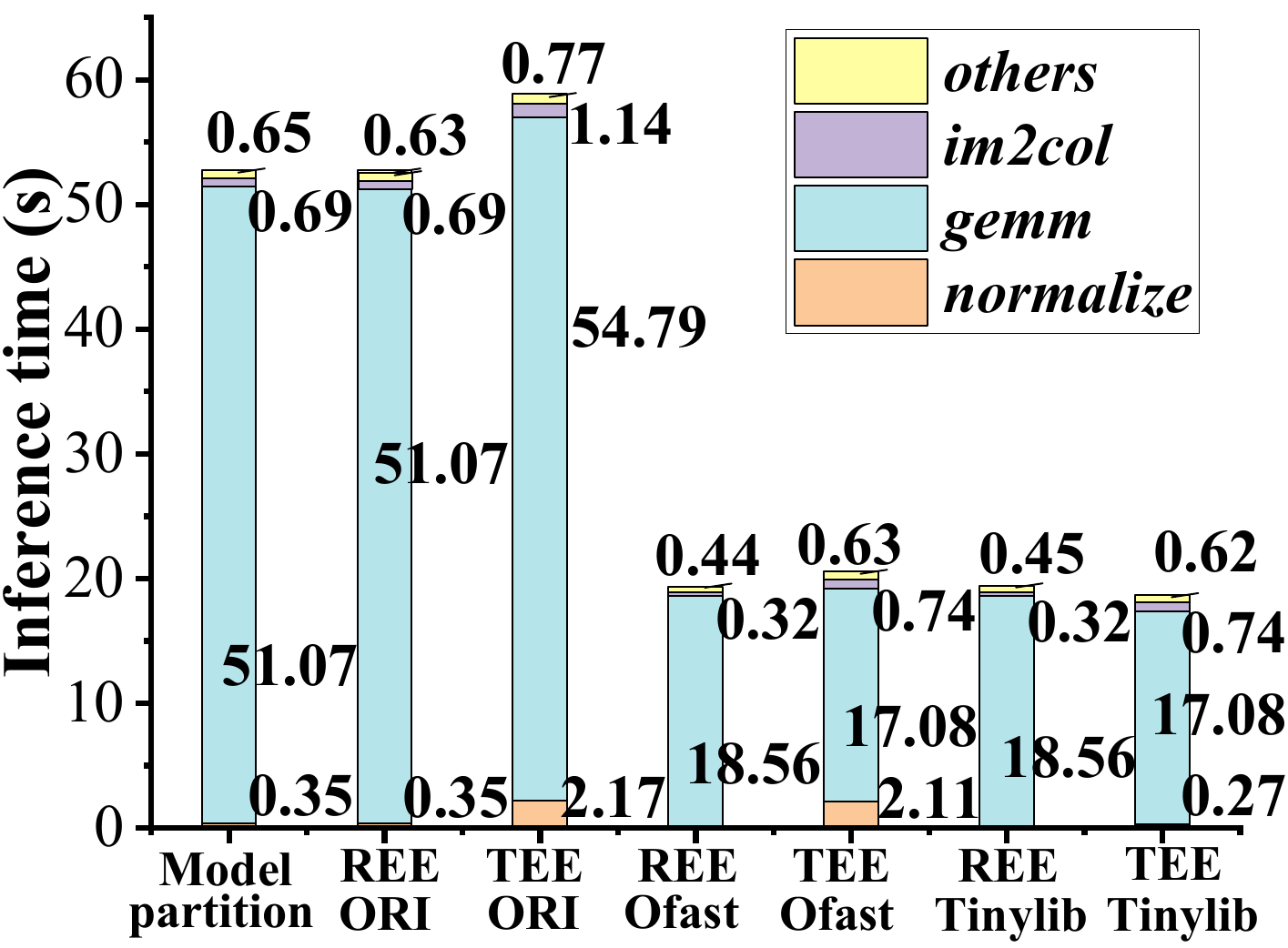}}
\end{minipage}\label{per-resnet}
\begin{minipage}[b]{0.26\linewidth}
\centering
\subfloat[][Darknet Reference]{\includegraphics[width=1\textwidth]{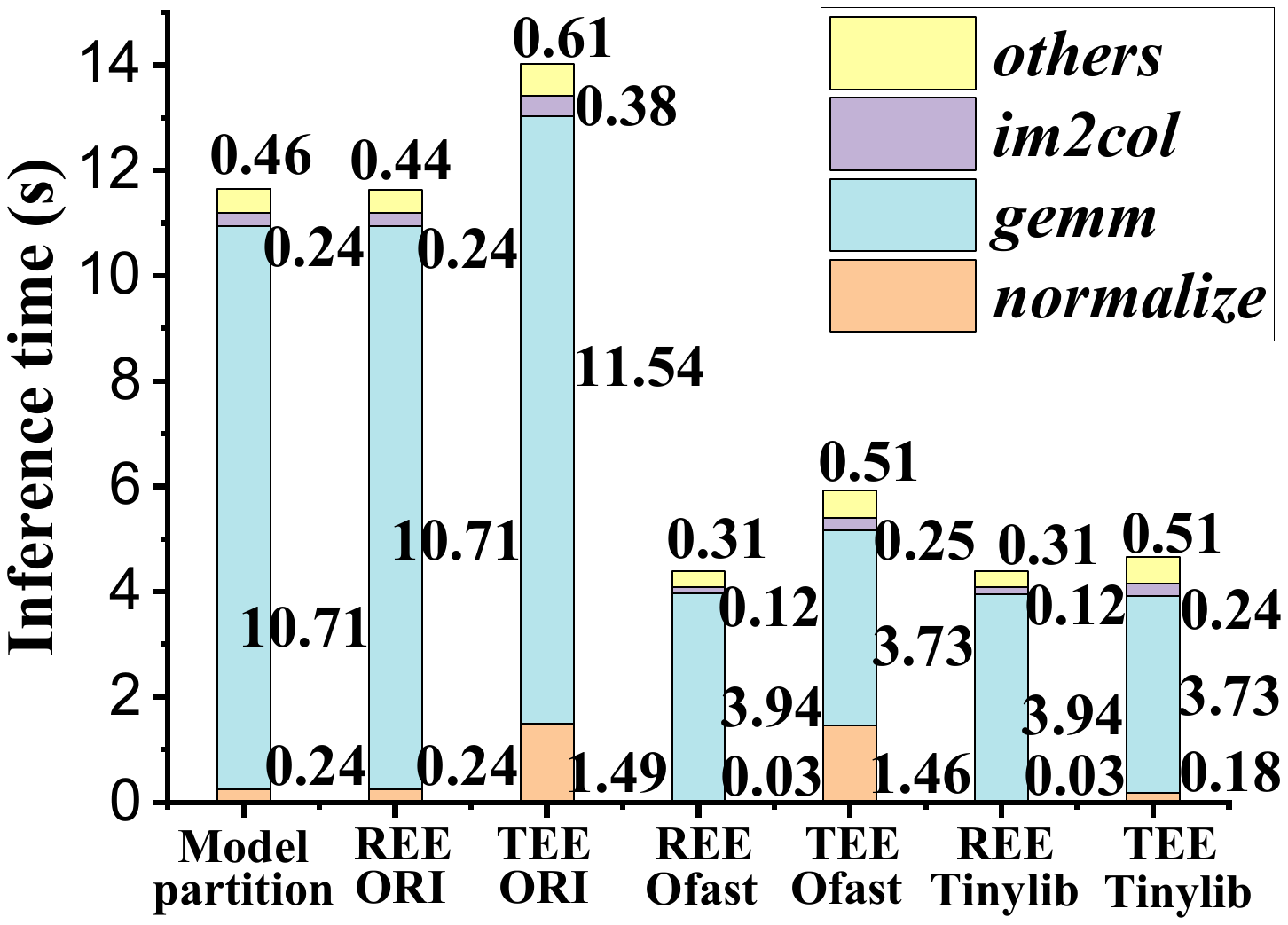}}
\end{minipage}
\begin{minipage}[b]{0.26\linewidth}
\centering
\subfloat[][Mobilenet]{\includegraphics[width=1\textwidth]{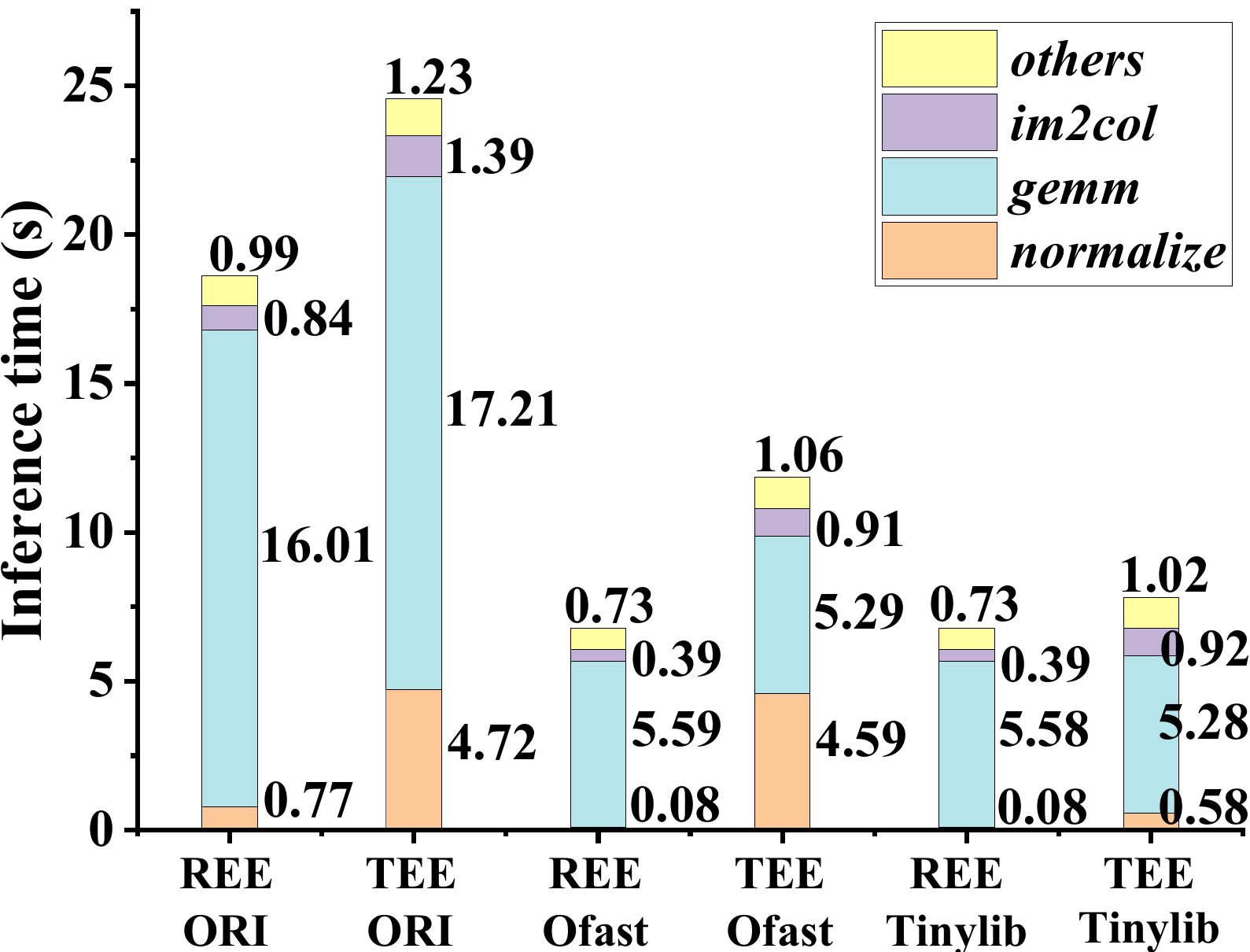}}
\end{minipage}
\caption{The performance on advanced compilation optimization: (a) Resnet-18; (b) Darknet Reference; (c) Mobilenet.}
\label{performance}
\end{figure*}

As shown in Fig. \ref{performance}, we tested the inference time of three DNN models, including the transmission of input image and prediction results, the network layer inference process, and the post-processing of prediction results, and also calculate the calculation time of \emph{normalize}, \emph{gemm},
\begin{figure*}[ht]
\centering
\begin{minipage}[b]{0.26\linewidth}
\centering
\subfloat[][Resnet-18]{\includegraphics[width=1\textwidth]{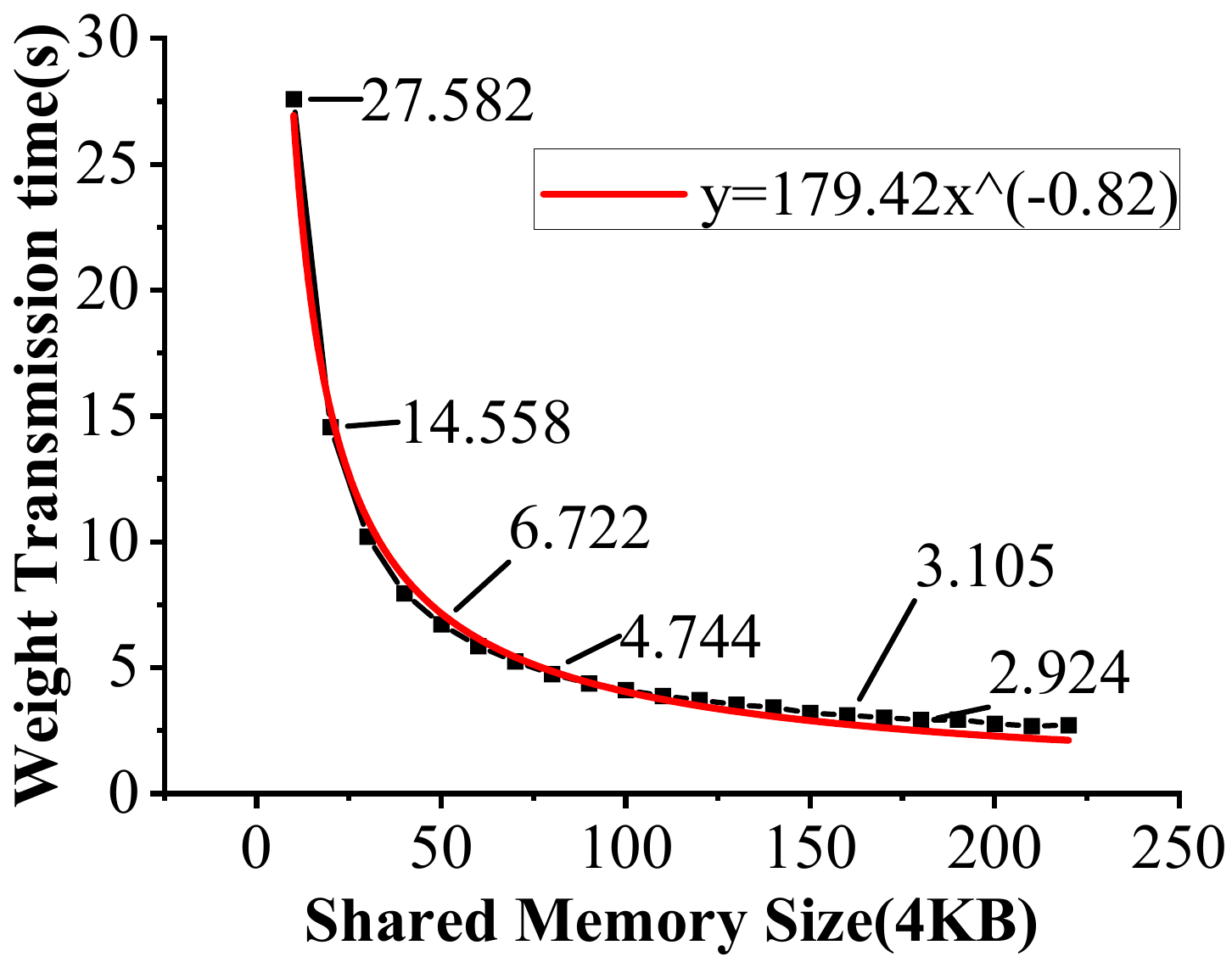}}
\end{minipage}
\begin{minipage}[b]{0.26\linewidth}
\centering
\subfloat[][Darknet Reference]{\includegraphics[width=1\textwidth]{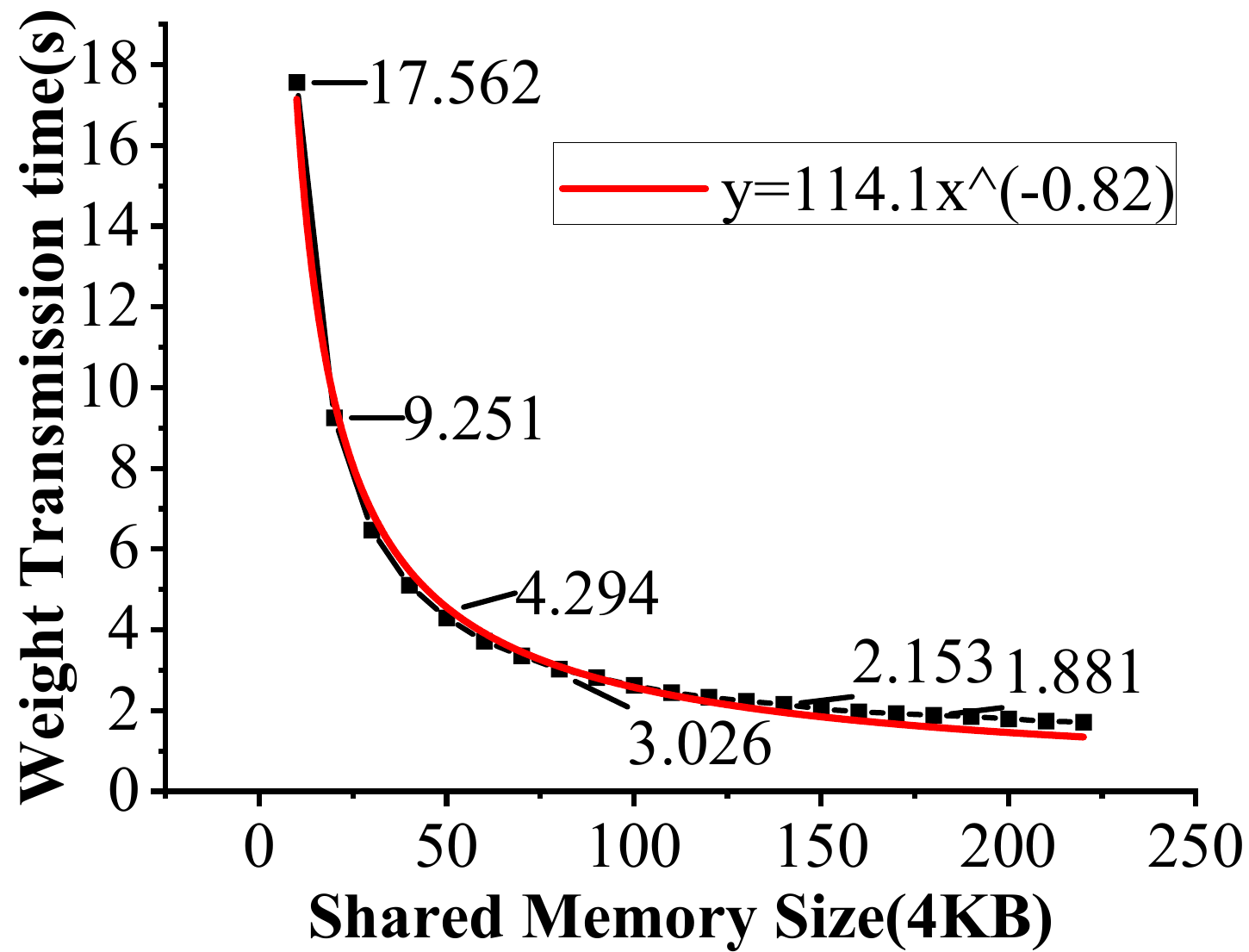}}
\end{minipage}
\begin{minipage}[b]{0.26\linewidth}
\centering
\subfloat[][Mobilenet]{\includegraphics[width=1\textwidth]{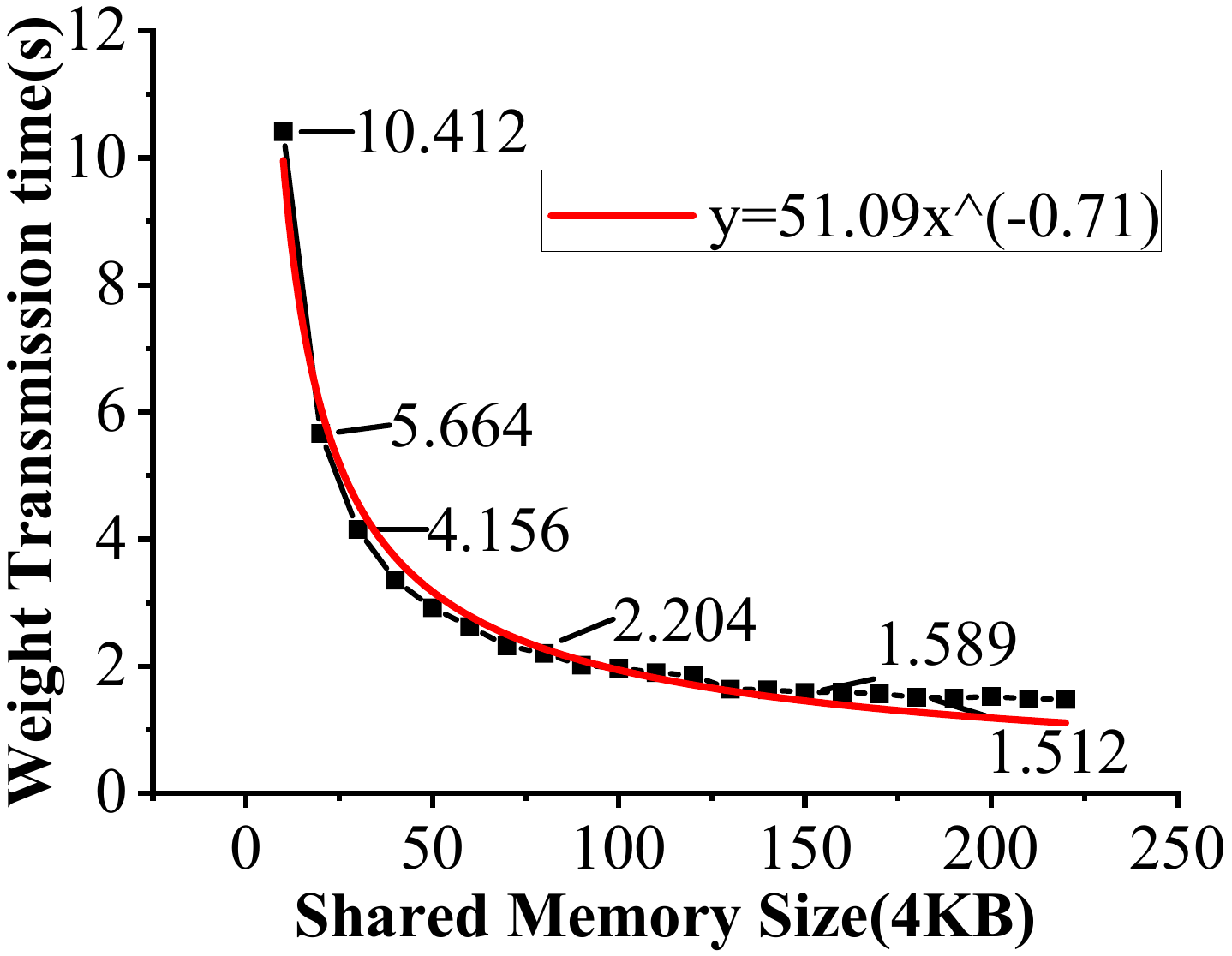}}
\end{minipage}
\caption{The optimal shared memory size fitting: (a) Resnet-18; (b) Darknet Reference; (c) Mobilenet. }
\label{weight}
\end{figure*}
\begin{figure}[ht]
  \centering
  \includegraphics[width=0.33\textwidth]{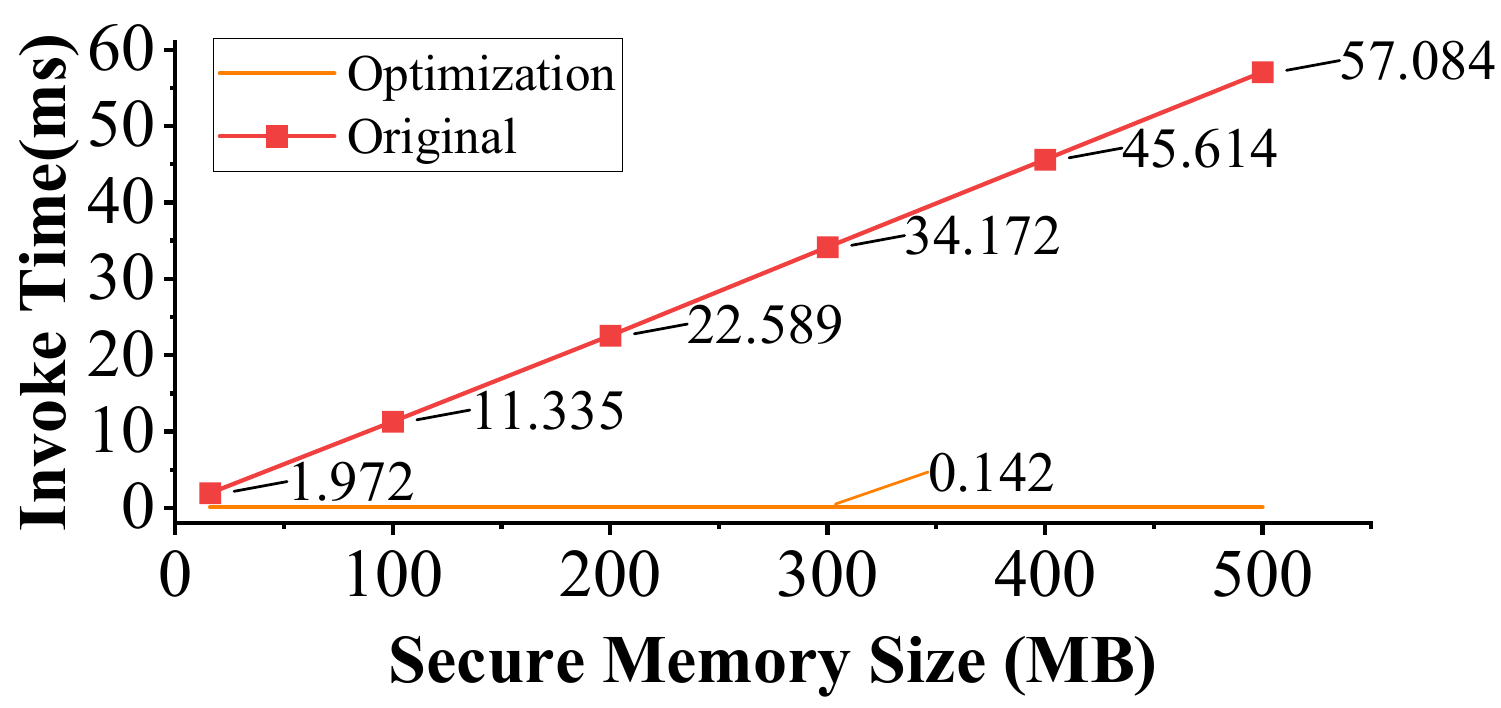}
  \caption{The time of \emph{invoke} operation in different secure memory sizes.}\label{invoketime}
\end{figure}
\begin{figure*}[ht]
\centering
\begin{minipage}[b]{0.26\linewidth}
\centering
\subfloat[][Resnet-18]{\includegraphics[width=1\textwidth]{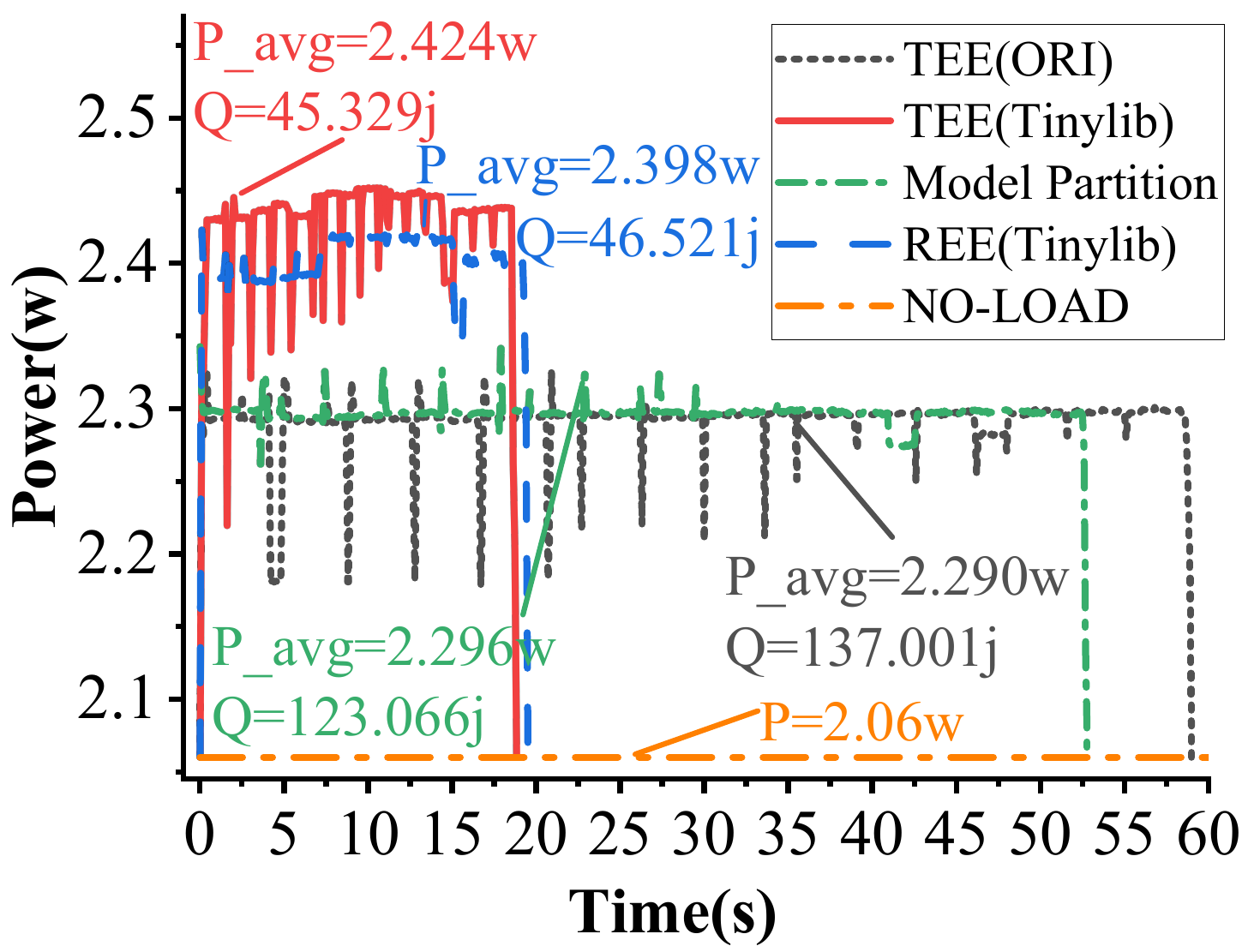}}
\end{minipage}
\begin{minipage}[b]{0.26\linewidth}
\centering
\subfloat[][Darknet Reference]{\includegraphics[width=1\textwidth]{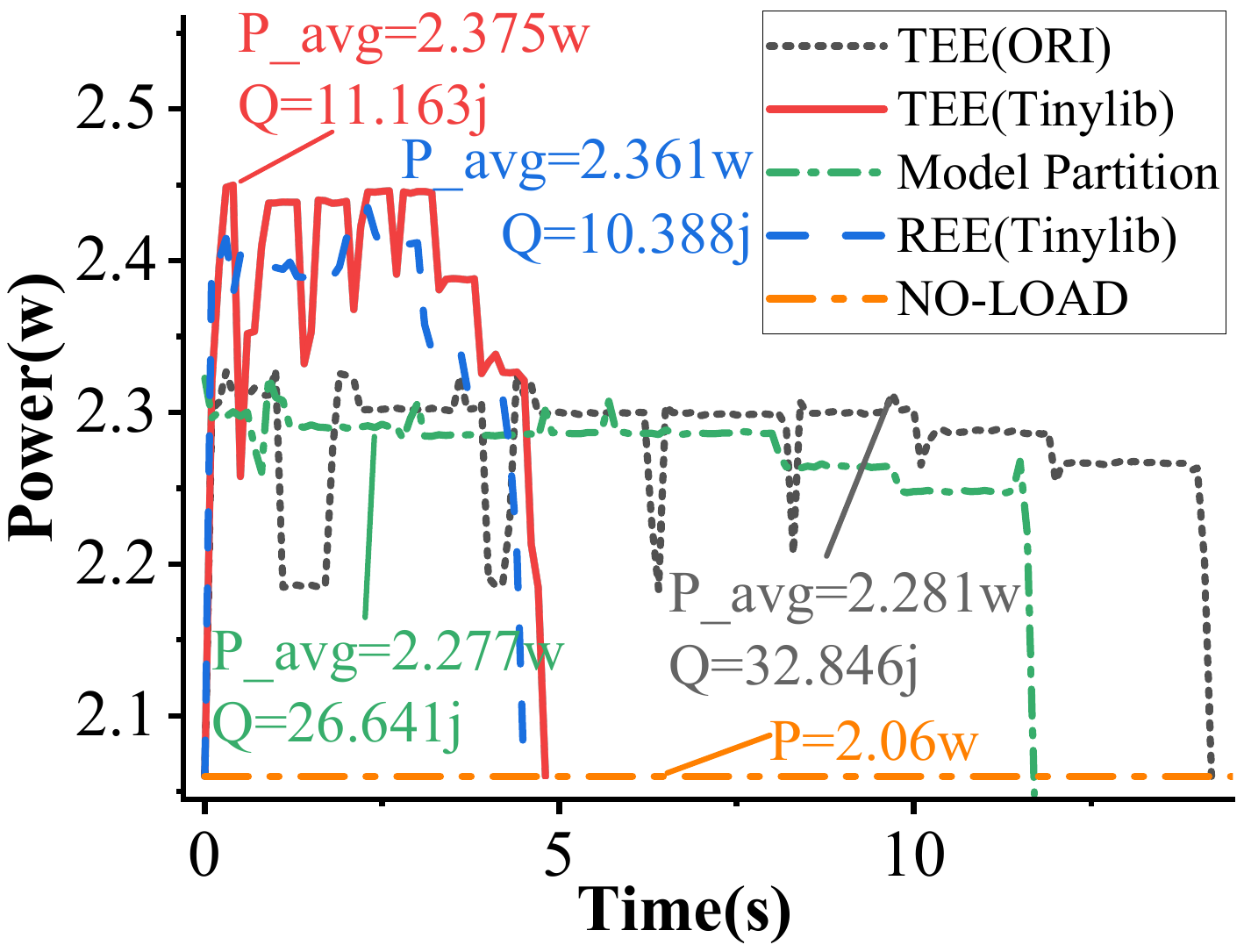}}
\end{minipage}
\begin{minipage}[b]{0.26\linewidth}
\centering
\subfloat[][Mobilenet]{\includegraphics[width=1\textwidth]{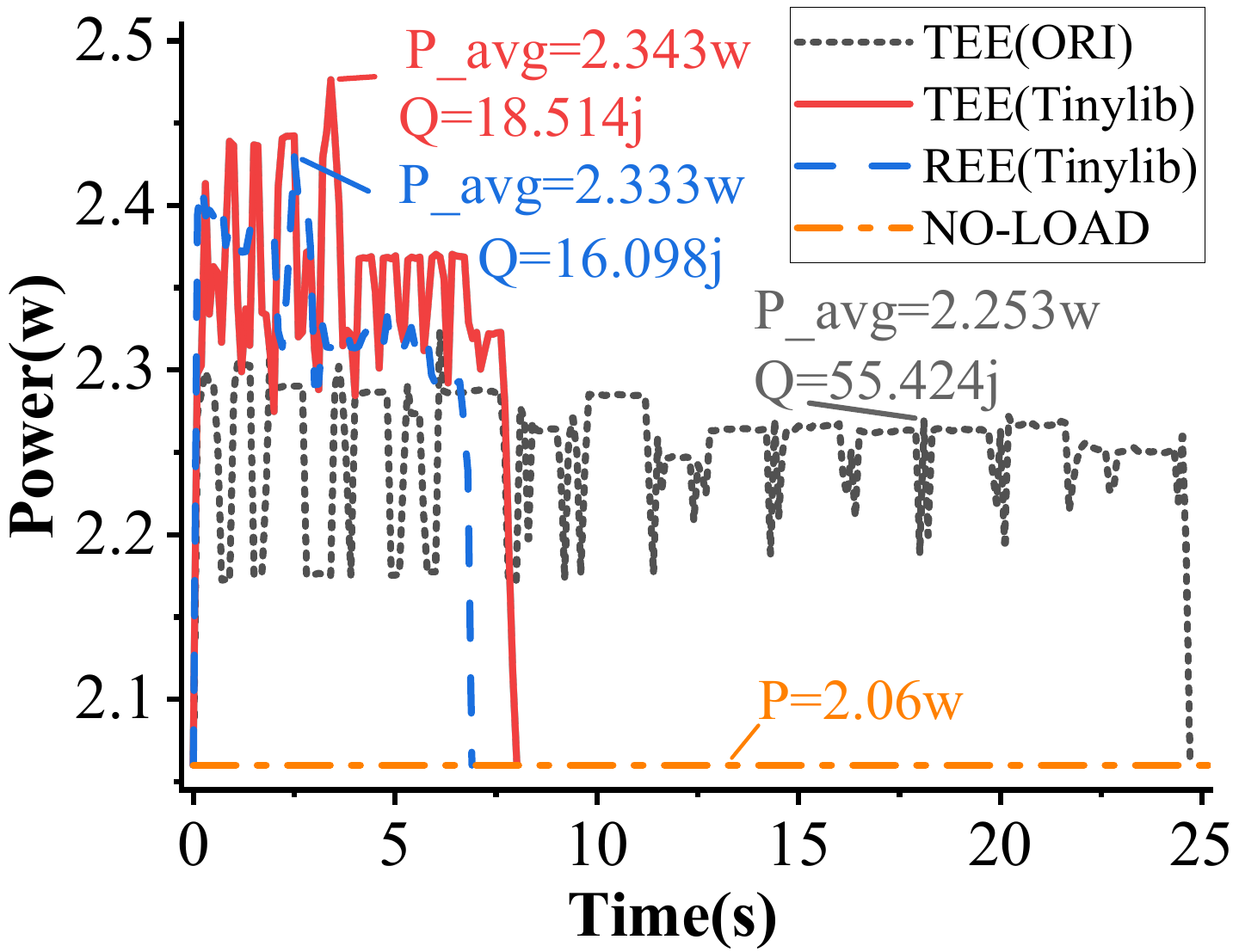}}
\end{minipage}
\caption{The performance on power consumption: (a) Resnet-18; (b) Darknet Reference; (c) Mobilenet. }
\label{power}
\end{figure*}
\emph{im2col} in the network layer inference process separately.
In the left of Fig. \ref{performance}, the results show that after using \emph{Ofast} compile optimization, the calculation time of \emph{im2col} and \emph{gemm} of Resnet-18 drops from $55.93$s to $17.82$s, almost $68.1$\% reduction in TEE. 
Similarly, the calculation time of Darknet Reference and Mobilenet decreases by $66.9$\% and $66.7$\% respectively.
When computing the \emph{convolution}, Tinylib first uses \emph{im2col} to convert the input three-dimensional data into a two-dimensional matrix, then use \emph{gemm} to speed up the \emph{convolution} calculation and the \emph{Ofast} compilation optimization to fully accelerate \emph{gemm} operations.
On the basis of using \emph{Ofast}, the time consumption of the normalizing process in TEE has also decreased significantly after using Tinylibm. The \emph{normalization} time of Resnet-18 decreases from $2.11$s to $0.27$s, almost $87.4$\% reduction in TEE. 
The \emph{normalization} time of Darknet Reference and Mobilenet also decreased by $87.5$\% and $87.4$\% respectively.
Because Tinylib uses the \emph{batch normalization} method for regularization, and \emph{sqrt} function is used for square root extraction when calculating norms. 
Combining the above results, TEE inference time is speeded by an average of $2.54$$\times$ (a range from $2.072$$\times$ to $2.863$$\times$) after \emph{Ofast} compilation optimization, and further speeded by an average of $3.13$$\times$ (a range from $3.011$$\times$ to $3.152$$\times$) after using Tinylibm.

We also compared with model partition solution, and use TEE to protect the last layer of the DNN model. 
Since previous work not supporting additional model extensions for Darknet,
we only counted the inference time of the Resnet-18 and Darknet Reference supported by Darknet. In the left and middle of Fig. \ref{performance}, the time taken by our design to perform Resnet-18 inference decreased from $52.764$s to $18.687$s, almost $64.6$\% reduction.
Darknet Reference also decreased by $59.9$\%.
The experimental results show that our solution has a higher performance than model partition solution.

\textbf{Memory Optimization.}
As shown in Fig. \ref{weight}, we set the shared memory size in $4$KB as the basic unit, considering that the current storage devices often use $4$KB alignment, and use the formula (3) to fit the relationship between the weight transfer time and the shared memory size. In the left of Fig. \ref{weight}, the fitting function of Resnet-18 is $y=179.42*x^{-0.82}$. 
 
The experimental results show that with the increase of shared memory, the weight transmission time is getting lower, but the decreasing trend is getting slower.  
We calculate the derivation of the fitting function to get the decreasing trend of weight transmission delay with the increase of shared memory.
According to our experience, we believe that when the derivative value is $-0.01$, the weight transfer delay and the shared memory size can be well balanced.
For example, we suggested that the most appropriate shared memory size for the Resnet-18 is $776$KB, and the corresponding transfer time is $2.387$s.

As shown in Fig. \ref{invoketime}, when the secure memory is increased from $16$MB to $500$MB, \emph{invoke} time increases from $1.972$ms to $57.084$ms, increasing approximately $29$$\times$. With the increase of secure memory, \emph{invoke} time increases approximately $11.437$ms for each $100$MB increase in secure memory. 
After optimizing the \emph{invoke} operation, the time for one \emph{invoke} decreases to $0.142$ms, and since there is no page table remap, the \emph{invoke} time does not increase with the increase of secure memory. The reduction is $92.8\%$ compared to 16 MB of secure memory and $99.8\%$ compared to $500$MB of secure memory.

\begin{table}[ht]
\caption{Comparisons with some state-of-the-art TEE-based Secure DNN inference Solutions.}\label{secure}
\begin{center}
\resizebox{\linewidth}{!}{
\begin{tabular}{|c|c|c|c|c|c|}
\hline
 & \emph{Arch} & \emph{Model} & \emph{Inference} & \emph{Cross-layer} & \emph{Network}  \\
\hline
LASAGANA\cite{li2022efficient} & Intel SGX  & Part & ALL &  \emph{Mid}  & Yes  \\
\hline
SCLERA\cite{kumar2022sclera} & Intel SGX & Part & Part & \emph{Mid}  & Yes \\
\hline
Occlumency\cite{lee2019occlumency} & Intel SGX & Public & ALL & \emph{High}  & Yes  \\
\hline
Plinius\cite{yuhala2021plinius} & Intel SGX & Part & Part & \emph{Low}  & Yes  \\
\hline
DarKnight\cite{hashemi2021darknight} & Intel SGX \& GPU & Part & Part &  \emph{High} &  No  \\
\hline
AegisDNN\cite{xiang2021aegisdnn} & Intel SGX \& GPU & Part & Part & \emph{Mid}  & No  \\
\hline
DarkneTZ\cite{mo2020darknetz} & ARM TrustZone & Public & Part & \emph{Low}  & No  \\
\hline
SecDeep\cite{liu2021secdeep} & ARM TrustZone & Part & Part & \emph{Mid}  & No \\
\hline
\textbf{Trusted-DNN\cite{liu2021trusted}} & ARM TrustZone & Part & ALL & \emph{High}  & No  \\
\hline
\textbf{Smart-Zone} & ARM TrustZone & ALL & ALL & \emph{Low} & No  \\
\hline
\end{tabular}}
\end{center}

\end{table}

\textbf{Power Consumption.}
As shown in Fig. \ref{power}, the no-load power is $2.06$w.
In the middle of Fig. \ref{power}, when running the Darknet Reference, the average power of all inference calculations using TEE protection is $2.375$w, and the power consumption by inference of an image is $11.163$j. Compared with REE, the average power of inference in TEE is increased by $0.6$\%, and the power consumption is increased by $7.5$\%. Because TEE requires additional \emph{invokes} to transmit the image and results compared to the inference in REE. The increase in calculation amount and additional transmission time lead to an increase in average power and power consumption. We also notice that the average power is increased from $2.281$w to $2.375$w, about $5$\% increment, compared with optimization not used in TEE. Because the hardware resources are fully utilized to achieve higher parallel computing after using the \emph{Ofast} compilation. The power consumption decreases from $32.846$j to $11.163$j, almost $66.0$\% reduction as the inference time after optimization is lower.
In the left and right subfigure of Fig. \ref{power}, the power consumption to perform inference on Mobilenet and Resnet-18 is similar to Darknet Reference.

We also compare with model partition solution. In the left of Fig. \ref{power}, the average power executing the Resnet-18 is increased from $2.296$w to $2.424$w, approximately $5.6$\% increment, and the power consumption is decreased from $123.066$j to $45.329$j, approximately $63.2$\% reduction.
Combined with the results of Darknet Reference in the middle of Fig. \ref{power}, the average power is increased by an average of $4.95$\% (a range from $4.3$\%
to $5.6$\%), and the power consumption is decreased by an average of $60.6$\% (a range from $58$\%
to $63.2$\%). The experimental results show that our entire protection can significantly reduce the power consumption of model inference in Smart-Zone.

\subsection{Comparing with SOTA Works}
In TABLE \ref{secure}, we compare with some SOTA works to highlight our unique contributions: \emph{Arch} means the TEE architecture; \emph{Model} means the protection of the pre-trained model; \emph{Inference} means the protection of the inference process; \emph{Cross-layer} means the number of world switches between TEE and REE, where \emph{Low} represents two world switches, \emph{Mid} represents a world switch close to the number of network layers, and \emph{High} represents a world switch far beyond the number of network layers; \emph{Network} means whether the needed to transmit data. AS the limited security memory size of TEEs, the above works must choose between security and performance, and difficult to protect valuable pre-trained models. Our work can allocate the best memory management strategy for each DNN model, while protecting the pre-trained model and inference process, with minimal overhead.

\section{Conclusion}
In this paper, we design and implement Tinylib, the first library that supports secure DNN inference for models trained by other famous frameworks and conveniently deploys on TrustZone-enabled consumer IoT devices for completely trusted isolation. Meanwhile, we reallocate an optimal secure memory size and shared memory size according to the requirements of pre-trained DNN models, and solve the memory overlapping conflict by priority assignment. We follow the TCB minimization guidelines and reuse some open-source libraries to minimize the size of S-Tinylib and Tinylibm. The evaluation based on real devices shows the effectiveness of our design and can significantly reduce inference time and energy consumption during model inference.
The authors have provided public access to their code and data at \href{}{https://github.com/nkicsl/SmartZone.git.}

\section*{Acknowledgment}
This work is partially supported by the National Natural Science Foundation (62272248), the Natural Science Foundation of Tianjin of China (21JCZDJC00740, 21JCYBJC00760, 23JCQNJC00010) and the Open Project Fund of State Key Laboratory of Computer Architecture, Institute of Computing Technology, Chinese Academy of Sciences (CARCH201905, CARCHA202108). 


\bibliographystyle{plain}
\bibliography{Reference}

\end{document}